\input harvmac
\overfullrule=0pt
\parskip=0pt plus 1pt
\sequentialequations
\def\adsf{\hbox{\rm adS}_5}
\def\adss{\hbox{\rm adS}_7}
\def\s5{S^5}
\def\sfour{S^4}
\def\half{{1\over 2}}
\def\del{\partial}

\def\np{Nucl.\ Phys.}
\def\pl{Phys.\ Lett.}
\def\prl{ Phys.\ Rev.\ Lett.}

\def\ap{Ann.\ Phys.}

\def\jhep{JHEP} 
\Title{\vbox{\rightline{BONN-TH/99-04}
\rightline{MRI-PHY/P990204}
\rightline{LMU-TPW-99-04}
\rightline{hep-th/9903042}}
}
{\vbox{\centerline{Stringy Corrections to the Wilson Loop}
\centerline{in ${\cal N}=4$ Super Yang-Mills Theory}}}
\medskip

\centerline{
Stefan F\"orste${}^1$,  Debashis Ghoshal${}^{2}$,
Stefan Theisen${}^3$}
\bigskip
\centerline{${}^1${\it Institute of Physics, Bonn University}}
\centerline{\it Nu\ss allee 12, Bonn 53115, Germany}
\smallskip
\centerline{${}^2${\it Metha Research Institute of Mathematics \&\
Mathematical Physics}}
\centerline{\it Chhatnag Road, Jhusi, Allahabad 211019, India}
\smallskip
\centerline{${}^3${\it Sektion Physik,
Universit\"at M\"unchen}}
\centerline{\it Theresienstrr\ss e 37, 80333 Munich, Germany}
\smallskip

\vglue .3cm
\bigskip

\noindent
We study stringy fluctuations as a source for corrections 
to the Wilson loop as obtained from the superstring on 
$\adsf\times\s5$/${\cal N}=4$ SYM correspondence. We give 
a formal expression in terms of determinants of 
two-dimensional operators for the leading order correction.

\Date{02/99}

\def\MALDADS{J.\ Maldacena, {\it The large N limit of superconformal 
field theories and supergravity}, Adv.\ Theor.\ Math.\ Phys.\ {\bf 2}
(1998) 231, {\tt hep-th/ 9711200}.}
\def\GUKLPO{S.\ Gubser, I.\ Klebanov and A.\ Polyakov, {\it Gauge
theory correlators from non-critical string theory}, \pl\ {\bf B428} (1998) 
105, {\tt hep-th/9802109}.}
\def\WIHOL{E.\ Witten, {\it Anti-de Sitter space and holography}, Adv.\ 
Theor.\ Math.\ Phys.\ {\bf 2} (1998) 253, {\tt hep-th/9802150}.}
\def\METTS{R.\ Metsaev and A.\ Tseytlin, {\it Type IIB superstring action in 
$\adsf\times\s5$ background}, \np\ {\bf B533} (1998) 109,
{\tt hep-th/9805028}.}
\def\MALLOOP{J.\ Maldacena, {\it Wilson loops in large N field theories}, 
\prl\ {\bf 80} (1998) 4859, {\tt hep-th/9803002}.}
\def\REYEE{S.-J.\ Rey and J.\ Yee, {\it Macroscopic strings as heavy quarks
in large N gauge theory and anti-de Sitter supergravity}, 
{\tt hep-th/9803001}. }
\def\KALLCOLL{R.\ Kallosh, J.\ Rahmfeld and A.\ Rajaraman, {\it Near horizon
superspace}, JHEP {\bf 10} (1998) 002, {\tt hep-th/9805217};
\hfill\break
R.\ Kallosh, {\it Superconformal action in Killing gauge},
{\tt hep-th/9807206};
\hfill\break
I.\ Pesando, {\it A kappa fixed type IIB superstring action on $\adsf
\times\s5$}, JHEP {\bf 11} (1998) 002, {\tt hep-th/9808020};\hfill
\break
R.\ Kallosh and J.\ Rahmfeld, {\it The GS string action on $\adsf\times\s5$},
\pl\ {\bf B443} (1998) 143, {\tt hep-th/9808038};\hfill\break
R.\ Kallosh and A.\ Tseytlin, {\it Simplifying superstring action on 
$\adsf\times\s5$}, JHEP {\bf 10} (1998) 016, {\tt hep-th/9808088};
\hfill\break
A.\ Rajaraman and M.\ Rozali, {\it On the quantization of the GS string
on $\adsf\times\s5$}, {\tt hep-th/9902046}. }
\def\AGFRMU{L.\ Alvarez-Gaum\' e, D.\ Freedman and S.\ Mukhi, {\it The 
background field method and the ultraviolet structure of the supersymmetric
nonlinear sigma models},  \ap\ {\bf 134} (1981) 85.}
\def\BAGR{T.\ Banks and M.\ Green, {\it Nonperturbative effects in 
$\adsf\times\s5$ string theory and $D=4$ susy Yang-Mills}, \jhep\ {\bf 5}:002
(1998), {\tt hep-th/9804170}. }
\def\GUKLTS{S.\ Gubser, I.\ Klebanov and A.\ Tseytlin, {\it Coupling constant
dependence in the thermodynamics of $N=4$ supersymmetric Yang-Mills
theory}, {\tt hep-th/9805156}.}
\def\PATH{J.\ Pawelczyk and S.\ Theisen, {\it $\adsf\times\s5$ metric at order
$\alpha'^3$}, {\tt hep-th/9808126}.}
\def\GILKEY{T.\ Branson, P.\ Gilkey and D.\ Vassilevich, {\it Vacuum
expectation value asymptotics for second order differential operators
on manifolds with boundary}, J.\ Math.\ Phys.\ {\bf 39} (1998) 1040,
{\tt hep-th/9702178};\hfill\break
P.\ Gilkey, {\it Invariance theory, the heat equation and the Atiyah-Singer
index theorem}, 2nd ed (1995), CRC Press.}
\def\GrSchWi{M.\ Green, J.\ Schwarz and E.\ Witten, {\it Superstring
theory}, vol.\ I, Cambridge University Press.}
\def\GROL{J.\ Greensite and P.\ Olesen, {\it Worldsheet fluctuations
and heavy quark potential in the adS/CFT approach}, {\tt hep-th/9901057}.}
\def\SNAIK{S.\ Naik, talk in the Workshop on {\it String theory, gauge 
theories and quantum gravity}, Puri, India (1998) (unpublished); and
private communications.}
\def\FORSTE{S.\ F\" orste, {\it Membrany corrections to the string 
anti-string potential in M5-brane theory}, {\tt hep-th/9902068}.}

{\nopagenumbers

\ftno=0

}

\newsec{Introduction}
The remarkable duality between type IIB string theory on 
$\adsf\times{\cal M}_5$, (${\cal M}_5$ a compact manifold), and conformally 
invariant supersymmetric Yang-Mills theory at the boundary of the anti-de 
Sitter space, proposed in \ref\MaldadS{\MALDADS} (and elaborated in
\ref\GuKlPo{\GUKLPO}\ref\WiHol{\WIHOL}), provides a new approach 
to the large $N$ limit of four-dimensional gauge theories. Many aspects 
of this non-trivial relation have been and indeed still are, the subject 
of numerous investigations. An example, that will be further studied 
in this letter, is the computation of the potential between a heavy 
quark-antiquark pair\ref\Malloop{\MALLOOP}\ref\ReYee{\REYEE}. 
In gauge theory this is conveniently done by evaluating the expectation 
value of the Wilson loop operator. In the adS/CFT scheme, one considers 
a string worldsheet with boundary fixed at the loop one is interested in. 
The (exponential of the) type IIB string action with this boundary 
condition is then the expectation value of the Wilson loop operator. 

In Refs.\Malloop\ReYee, the authors considered the case ${\cal M}_5=S^5$
leading to the maximally supersymmetric ${\cal N}=4$ Yang-Mills theory.
The expectation value of the Wilson loop was calculated to the lowest order
by evaluating the area of the string worldsheet. A macroscopic string
is stretched between the quark and antiquark at the boundary of the
anti-de Sitter space. The non-trivial metric of this space means that it
is energetically favourable for the string to fall in the interior of the
adS space along a geodesic connecting the two points on the boundary.
The string action is the area of the worldsheet in the induced metric. 

The full superstring action in the background $\adsf\times\s5$ has
meanwhile been constructed in \ref\MetTs{\METTS}, and further
considered in \ref\KallColl{\KALLCOLL}. This enables one to address 
the question of stringy corrections to the classical area of the worldsheet.
In this letter we will consider the sigma model corrections at one loop by 
expanding the string action to quadratic order in fluctuation. We first 
recapitulate briefly the set up and calculation in \Malloop\ReYee. We
then start with the superstring action in \MetTs\ and expand it around
the classical background using the normal coordinates\ref\AGFrMu{\AGFRMU}.
The issue of gauge fixing to actually evaluate the determinant is
discussed next. We finally give the result in terms of determinants of
two-dimensional second order differential operators and discuss the
issue of UV divergences. 

\newsec{Wilson loop in adS/CFT correspondence}
Let us consider a static configuration of a quark-antiquark pair 
separated by a distance $L$. The Wilson loop is a rectangular one
the sides of which are parallel to the time and one of the space 
directions. The length of the temporal side $T$ is taken to infinity. 
In the dual string description there is a macroscopic string with
endpoints fixed on the quarks by Dirichlet boundary condition. If,
for simplicity we assume that the string ends are at the same point
on $S^5$, the minimum energy configuration is that for which the
string is stretched along a geodesic in $\adsf$. This follows from the
fact that the classical string action (with the overall factor $\alpha'$
set to 1) is
\eqn\nambugoto{
S_{NG} ={1\over 2\pi}
 \int  d^2\sigma\sqrt{-\det ||G_{MN}\del_i x^M\del_j x^N||}, }
where the $G_{MN}$ is the metric on $\adsf\times\s5$ given by
\eqn\adsmetric{
ds^2 =
R^2\left[ u^2\left(-(dx^0)^2 + (dx^1)^2 +(dx^2)^2+(dx^3)^2\right) +
{du^2\over u^2} + d\Omega_5^2\right]. }
The spaces $\adsf$ and $\s5$ have the same value of radius 
$(4\pi gN)^{1/4}$. Note that we have rescaled $u\to R^2u$ compared to 
\Malloop\ to make the sigma model loop counting parameter 
$R^2/\sqrt{\alpha'}$ manifest.

It is convenient to adopt a `static gauge' to describe the macroscopic 
string. Let $X^M$ denote the classical values of the string coordinates.
We set $X^0=\tau$, $X^1=\sigma$ and assume that all the other coordinates  
except $U=U(\sigma)$ are independent of the worldsheet coordinates
$(\tau,\sigma)$. The radial coordinate $U$  has a nontrivial dependence 
on $\sigma$ only, which is implicitly given by\Malloop
\eqn\usigma{
\del_\sigma U = \pm {U^2\over U_0^2}\sqrt{U^4-U_0^4} }
with a constant $U_0 = (2\pi)^{3/2}/\Gamma(1/4)^2 L$. ($U_0$ is the 
closest the string comes to the origin of the $\adsf$ space, and once
again we have rescaled it by $R^2$.) Due to the symmetry of the string
configuration it will be sufficient to restrict to the range 
$0\leq \sigma \leq L/2$ and take the postive root in \usigma.

With this choice of classical solution, the metric induced on the 
worldsheet $h^{cl}_{ij}$ is given by
\eqn\indmetric{
{ds^2_{cl}\over R^2 } = - U^2\; d\tau^2 + {U^6\over U_0^4}\;d\sigma^2.}
The energy of the quark pair is calculated by evaluating the area 
$A_R$ of the string worldsheet using this induced metric, and is (after 
subtracting an infinite contribution of the quark mass\Malloop)
\eqn\clenergy{
E = -\, {4\pi^2\sqrt{2g_{YM}^2N}\over\Gamma(1/4)^4 L}, }
that is of the form of Coulomb law. The $1/L$ dependence of the energy 
is a consequence of conformal invariance, but the $g_{YM}^2N=gN$, is 
non-perturbative from the point of view of the gauge theory.

How is this result calculated in classical string theory, corrected? One
possible source of correction from the change in geometry has been
ruled out by a number of authors\ref\BaGr{\BAGR}\MetTs\KallColl\ who
argued that the $\adsf\times\s5$ is an exact string background\foot{This
is no longer true for gauge theory at finite 
temperature\ref\GuKlTs{\GUKLTS}\ref\PaTh{\PATH}.}. However, 
corrections can arise from taking into account the fluctuation of 
the string around the given classical configuration. For this we
will need to start with the type IIB superstring action in the 
$\adsf\times\s5$ background. Let us briefly sketch our approach
before we go into the details. 

We are going to replace the classical saddle-point approximation of 
\Malloop\ by a functional integral over the fluctuations
\eqn\correction{
W(C) = \int [{\cal D}\delta X] [{\cal D}\delta \theta]\;
e^{-S_{IIB}\left(X+\delta X, \delta\theta\right)},}
where $\delta X, \delta \theta$ denote quantum fluctuations of the
bosonic and fermionic coordinates. It was observed in \MetTs\ 
is that the string action on $\adsf\times\s5$ with radii $\mp R$ is 
$R^2$ times the action on $\adsf\times\s5$ with unit radii. 
Therefore as expected $1/R^2$ plays the role of a loop expansion 
parameter. The superstring action of \MetTs\ in this background is
a Green-Schwarz type action and is invariant under worldsheet 
diffeomorphism and a local fermionic kappa symmetry. Further there is
no supersymmetry on the worldsheet. However we will find that, much as 
in the case of flat space, a gauge fixing condition for kappa symmetry 
makes worldsheet fermions out of the fermionic coordinates of the 
target space. 


\newsec{Superstring action in $\adsf\times\s5$ background}
In Ref.\MetTs, Metsaev and Tseytlin have constructed an 
action for the IIB theory in the $\adsf\times\s5$ background. The action 
is defined as a covariant $\kappa$-symmetric two-dimensional 
sigma model on a supercoset appropriate for this background.
Explicitly, the part of the Lagrangian relevant to us here 
is
\eqn\action{
\eqalign{
{\cal L} &= - \half\sqrt{-h} h^{ij}\left(e^{\hat a}_i - i\bar\theta^I
\hat{\gamma}^{\hat a}(D_i\theta)^I\right)\left(e^{\hat a}_j - 
i\bar\theta^J\hat{\gamma}^{\hat a}(D_j\theta)^J\right)\cr
&\qquad -i\epsilon^{ij}e^{\hat a}_i\left(\bar\theta^1
\gamma^{\hat a}(D_j\theta)^1 - 
\bar\theta^2\gamma^{\hat a}(D_j\theta)^2\right), }}
where the notation used is same as in \MetTs. 
In particular $h_{ij}$ is a two-dimensional metric, 
$e^{\hat a}_i = \left(e^a_\mu\del_i x^\mu,
e^{a'}_{\mu'}\del_i y^{\mu'}\right)$ is defined in terms of vielbeins 
of $\adsf$ and $\s5$, 
$\hat{\gamma}^{\hat a}=\left(\gamma^a,i\gamma^{a'}\right)$
satisfy the $SO(4,1)$ and $SO(5)$ Clifford algebras respectively and
$\theta^I, I=1,2$ label the two sets of spinors of
these. The two-dimensional indices $i,j$ run over 1 and 2, 
$(a,a')=(0,\cdots,4;5,\cdots 9)$ are (flat) tangent space indices for 
$\adsf\times\s5$, and similarly for the curved indices $(\mu,\mu')$. 

In order to compute the one loop contribution we will expand
\action\ to second order in the fluctuations around the classical 
solution of \Malloop. In this background the metric on the worldsheet
$h_{ij}$ is the one induced from the target space \indmetric, and we
will fix $h_{ij}$ to this value, (which we will call $h^{cl}_{ij}$),
by exploiting worldsheet diffeomorphism. Notice that this differs from 
the standard practice of working with a flat (or conformally flat) 
worldsheet metric. 

The classical solution is non-trivial only in the bosonic part along
the $\adsf$ directions. Consequently fluctuations in bosonic degrees
of freedom along the $\adsf$ and $\s5$ space, and fluctuations in the
fermionic variables all decouple (to second order). Therefore one can
study these independently and add up their contributions.

\newsec{Fluctuations of the bosonic coordinates}
\subsec{The $\adsf$ part}
The metric in the $\adsf$ space is
\eqn\adsfivemet{
ds_{\adsf}^2 = u^2dx^a dx^b\eta_{ab} + {du^2\over u^2}.}
In the above $a,b=0,\cdots,3$; however, in the following the label
$a=4$ will refer to the `radial' coordinate $u$, and its quantum 
fluctuations. The classical solution has been reviewed in section 2.
We now expand the action in terms of the normal coordinates using
standard technology\AGFrMu. To quadratic order this leads to
\eqn\fluctwo{
\eqalign{
{\cal L}_{\adsf} = &- \sqrt{-h_{cl}}\Bigg[ 1 + \half 
h_{cl}^{ij}\eta_{ab}D_i\xi^a D_j\xi^b + \eta_{ab}\xi^a\xi^b\cr
& \qquad\qquad\qquad - \half h_{cl}^{ij}G_{\mu\lambda}G_{\nu\rho}
(\del_i X^\mu)(\del_j X^\rho) E^\lambda_a E^\nu_b\xi^a\xi^b\Bigg],\cr}}
where, $D_i\xi^a = \del_i\xi^a + (\del_i X^\mu)\omega^a_{\mu b}\xi^b$. 
In the above we have used the expression of the Riemann tensor in
terms of the metric
%
$R_{\mu\nu\rho\lambda} = - \left(G_{\mu\rho}G_{\nu\lambda}
- G_{\mu\lambda}G_{\nu\rho}\right)$, 
%
in adS space. 

However eqn \fluctwo\ is not the end of the story, as the variation of
the metric $h_{cl}^{ij}$ lead to nontrivial constraints
\eqn\constraints{
{\cal C}_{ij}\equiv E_{ij}(\xi) - \half h^{cl}_{ij}
\left(h_{cl}^{kl}E_{kl}(\xi)\right) \approx 0, }
where
%
$E_{ij}(\xi) = G_{\mu\nu}(x_{cl})\left(\del_ix_{cl}^\mu E^\nu_aD_j\xi^a + 
\del_jx_{cl}^\nu E^\mu_aD_i\xi^a\right)$. 
%
These constraints need to be taken into account. This can be done by 
following standard procedure. However it turns out to be easier to 
calculate the full induced metric (including fluctuations), which can 
then be shown to be equal to the sum of \fluctwo\ plus Lagrange 
multipliers times the constraints \constraints. 

In order to write the final form of the $\adsf$ part of the lagrangian,
let us introduce the following linear combinations
\eqn\defnofxi{
\eqalign{
\xi^\parallel &= {U_0^2\over U^2}\xi^1 + 
{\sqrt{U^4-U_0^4}\over U^2}\xi^4\cr
\xi^\perp &= -{\sqrt{U^4-U_0^4}\over U^2}\xi^1+
{U_0^2\over U^2}\xi^4.\cr }}
These are the (normalized) fluctuations along the direction parallel
(respectively perpendicular) to the classical string configuration. 
(That these indeed parametrize the fluctuations parallel and 
perpendicular to the string background is most evident in terms of the 
normal coordinates $\xi^\mu$ with curved indices.) In terms of these 
variables the quadratic part of the lagrangian takes the following 
simple form
\eqn\finaladslag{\eqalign{
{\cal L}^{(2)}_{\adsf} = & -\half\sqrt{-h_{cl}}\Bigg[h_{cl}^{ij}\left(
\del_i\xi^\perp\del_j\xi^\perp + \del_i\xi^2\del_j\xi^2 + 
\del_i\xi^3\del_j\xi^3\right)\cr
&\qquad\qquad\qquad +2\left(1-{U_0^4\over U^4}\right)
\left(\xi^\perp\right)^2 + 2\left(\xi^2\right)^2 + 
2\left(\xi^3\right)^2\Bigg]. }}
In writing \finaladslag, we have ignored some total derivative terms. 
One observes that the fluctuations $\xi^0$ and $\xi^\parallel$ along
the worldsheet have dropped out of the action. This is a consequence
of the worldsheet diffeomorphism, which is completely fixed if 
we eliminate two redundant degrees of freedom by choosing
\eqn\bosefix{
\xi^0 = \xi^\parallel = 0 .}
This gauge choice is analogous to the non-covariant light-cone gauge,
and is consistent with the static gauge employed to write the classical
solution. 
 
Finally we notice that the covariant laplacian of the induced metric 
\indmetric\ with its canonical connection,
$\Delta_{cl} = {1\over\sqrt{h_{cl}}}\del_i\left(
\sqrt{h_{cl}}h_{cl}^{ij}\del_j\right)$, appears in \finaladslag. For
future use, let us rewrite \finaladslag\ as
\eqn\reallyfinal{
{\cal L}^{(2)}_{\adsf}=\half\sqrt{h_{cl}}\left[\sum_{a=2,3,\perp}
\xi^a\Delta_{cl}\xi^a - 2\left(\xi^2\right)^2 - 2\left(\xi^3\right)^2
+ (R^{(2)}-4)\left(\xi^\perp\right)^2\right], }
where, $R^{(2)}=2(U^4+U_0^4)/U^4$ is the scalar curvature of the 
two-dimensional induced metric. The fluctuations $\xi^2$ and $\xi^3$ in 
the transverse directions are seen to be massive, while $\xi^\perp$
moves in a potential. In addition, as $U\to\infty$ the fluctuations 
must be required to vanish as that is where the heavy quarks sit. 

\subsec{The $\s5$ part}
We have assumed for simplicity a trivial background for the $\s5$ part. 
Both the quarks are at the same point on $\s5$, and this classical 
position is independent of the worldsheet coordinates $(\tau,\sigma)$. 
Let $\eta^{a'}$, $a'=5,\cdots,9$ be the normal coordinates denoting
the quantum fluctuations on the sphere. These variables behave like the
$\xi^{2,3}$ fluctuations in the $\adsf$ space. The lagrangian relevant 
for the $\s5$ part is
\eqn\sfpiece{
{\cal L}^{(2)}_{\s5}=-\half\sqrt{-h_{cl}}\;
h_{cl}^{ij}\del_i\eta^{a'}\del_j\eta^{a'} 
= \half\sqrt{h_{cl}}\; \eta^{a'}\Delta_{cl}\eta^{a'}.}
The fluctuations are massless and the the second order operator is just
the laplacian of the induced metric.

\newsec{Fluctuations of the fermionic coordinates}
Let us start by recalling the covariant derivative $(D_j\theta)^I$ 
appearing in \action:
\eqn\covder{
\eqalign{
(D_j\theta)^I = D_j^{IJ}\theta^J &= \left[\delta^{IJ}\left(\del_j + {1\over 4}
(\del_j x_{cl}^\mu)\omega_\mu^{ab}\gamma^{ab}\right) - {i\over 2}\epsilon^{IJ}
(\del_jx_{cl}^\mu)e_\mu^a\gamma^a\right]\theta^J\cr
&= {\cal D}_j\theta^I - {i\over 2}\epsilon^{IJ}
(\del_jx_{cl}^\mu)e_\mu^a\gamma^a\theta^J.\cr  }}
In \MetTs\ there are additional terms in the above definition, but in our
context those vanish. The first term ${\cal D}_j\theta^I = \del_j\theta^I + 
{1\over 4}(\del_j x_{cl}^\mu)\omega_\mu^{bc}\gamma^{bc}\theta^I$, is the 
standard covariant derivative on the fermions. The additional second
term appears due to the non-trivial coupling to the RR 5-form field
strength. Substituting above in \action, and using the properties of
gamma matrices and the fermions given in \MetTs, the fermionic part of
the action in the background of the macroscopic string is compactly 
written as
\eqn\fermilag{
{\cal L}_F = -\sqrt{-h_{cl}}
\left(\matrix{\bar\theta^1 &\bar\theta^2}\right)\left(\matrix{
2ie^a_\mu(\del_i x_{cl})^\mu\gamma^a{\cal P}_-^{ij}{\cal D}_j 
&1-{\cal B}\cr 
-1-{\cal B} 
& 2ie^a_\mu(\del_i x_{cl})^\mu\gamma^a{\cal P}_+^{ij}{\cal D}_j }\right)
\left(\matrix{\theta^1\cr \theta^2}\right) ,}
where ${\cal B}={1\over 2\sqrt{-h_{cl}}}\epsilon^{ij}e^a_\mu e^b_\nu
(\del_i X^\mu)(\del_j X^\nu)\gamma^{ab}$, and 
${\cal P}^{ij}_\pm=\half\left(h^{ij}_{cl}\pm \epsilon^{ij}/
\sqrt{-h_{cl}}\right)$ are projection operators similar to the ones 
in flat space\ref\GSW{\GrSchWi}.

Let us define the following combination of gamma matrices
\eqn\gammapape{\eqalign{
\gamma^\parallel &= {U_0 ^2\over U^2}\gamma^1 + 
{\sqrt{U^4-U_0^4}\over U^2}\gamma^4\cr
\gamma^\perp &= -{\sqrt{U^4-U_0^4}\over U^2}\gamma^1 + 
{U_0 ^2\over U^2}\gamma^4,\cr}}
in analogy with \defnofxi, and let $\gamma^\pm=\half\left(
\gamma^0\pm\gamma^\parallel\right)$. In the background of the 
classical solution, the diagonal terms are simply 
$i\gamma^\pm\left({1\over U}{\cal D}_\tau \pm 
{U_0^2\over U^3}{\cal D}_\sigma\right)$, and $\bar\theta^1{\cal B}\theta^2
=\bar\theta^2{\cal B}\theta^1 = -(2U^4/U_0^4)\bar\theta^1
\gamma^{0\parallel}\theta^2$. 

Now recall that the action \action\ or
\fermilag\ has a local fermionic gauge symmetry, the so called 
$\kappa$-symmetry, which has to be fixed so as to remove the redundant 
fermionic degrees of freedom. It turns out that a most convenient choice 
is to set
\eqn\kappafix{
\gamma^-\theta^1 = 0 \qquad ,\qquad \gamma^+\theta^2 = 0 .}
With this choice, the lagrangian \fermilag\ simplifies to
\eqn\expferlag{
{\cal L}_F =-\sqrt{-h_{cl}} 
\left(\matrix{\bar\theta^1 &\bar\theta^2}\right)\left(
\matrix{i\gamma^+ \bar D_+ &2\cr
-2 & i\gamma^-\bar D_- }\right)
\left(\matrix{\theta^1\cr \theta^2}\right) ,}
where we have defined the two-dimensional covariant derivative
with tangent space indices
\eqn\dpm{
\bar D_{\pm} = {1\over U}{\cal D}_\tau\pm {U_0^2\over U^3}{\cal D}_\sigma 
= \varepsilon^\tau_0 D_\tau\pm \varepsilon^\sigma_1 D_\sigma, }
$\varepsilon_0^\tau$ and $\varepsilon_1^\sigma$ being a set of
two-dimensional (inverse) vielbeine of the classical induced 
metric \indmetric.

The equations of motion that follow from this Lagrangian are
\eqn\fermieom{\eqalign{
\gamma^+\left(\del_+ + {\sqrt{U^4-U_0^4}\over 2U^2}\right)\theta^1 
+ \theta^2 & = 0\cr
\gamma^-\left(\del_- - {\sqrt{U^4-U_0^4}\over 2U^2}\right)\theta^2 
- \theta^1 & = 0,\cr }}
where the derivatives are with respect to tangent space indices on
the worldsheet, and their definitions are similar to \dpm\ above.
However the above form is somewhat deceptive as $\gamma^\pm$ depend 
on $\sigma$, and hence are not covariantly constant. This situation 
is remedied by exploiting the $\kappa$-symmetry fixing condition 
\kappafix. After some straightforward manipulations, one arrives 
at
\eqn\fermieomtwo{\eqalign{
i\gamma^0\nabla_+\theta^1+\theta^2\equiv 
i\gamma^0\left( \del_+ +{\omega \over 2}+A\right)\theta^1 +
\theta^2 &= 0\cr
i\gamma^0\nabla_-\theta^2 -\theta^1\equiv
i\gamma^0\left( \del_- -{\omega \over 2}-A\right)\theta^2 -
\theta^1 &= 0, \cr }}
where $\omega=\varepsilon^\tau_0\omega_\tau^{01}$ is the contribution
from the spin connection and $A={U_0^2\over U^2}\gamma^{14}$ is an
additional gauge connection. 

Now with the help of the following definition for two-dimensional 
gamma-matrices
\eqn\twodgamma{
\rho^+ = \left(\matrix{ 0 & 0\cr \gamma^0 &0 }\right)\quad ,\quad
\rho^- =\left( \matrix{0 & \gamma^0\cr 0 & 0}\right) }
%
the equations of motion are compactly expressed as
\eqn\eomcovariant{
\left( i\rho^j\nabla_j +\rho^3\right)\left(\matrix{\theta^1\cr
\theta^2}\right) = 0 ,}
with $\rho^3=$diag({\bf 1},$-${\bf 1}), and 
$\theta=\left(\matrix{\theta^1\cr \theta^2}\right)$ is a `two
component' spinor of the two-dimensional worldsheet. This is 
the adS analogue of the well-known `metamorphosis' of target-space 
spinors into world-sheet spinors\GSW. 

Coming back to \fermieomtwo\ now, we see that $\theta^1$ (say) is
completely determined in terms of $\theta^2$, which should be treated
as independent fermionic fields. Since each $\theta$ had, to start
with, 16 components and the gauge condition fixing $\kappa$-symmetry
\kappafix\ reduces these by half, the fermions have altogether eight
on-shell degrees of freedom. This of course matches with those of 
the bosonic fields. 

The coupled set of first order equations \fermieomtwo\ can be traded
for the second order equation
\eqn\fermisecond{
\left(-\Delta_F - {1\over 4}R^{(2)} + 1\right)\theta^2 = 0, }
for $\theta^2$ alone, and similarly for $\theta^1$. In the above,
the $4\times 4$ matrix operator $\Delta_F=h^{ij}\nabla_i\nabla_j$ 
is the laplacian of the generalized covariant derivative (including 
gauge connection) \fermieomtwo\ acting on the fermions.

\newsec{Towards evaluation of the determinants}
%
We are now in a position to perform the functional integration over
the fluctuations, and give a formal expression for the one-loop result.
Collecting \reallyfinal, \sfpiece\ and \fermisecond, we find the
following expression for $W(C)$ in \correction: 
\eqn\effa{
W(C) = e^{-A_R}\;{\det\left(-\Delta_F - {1\over 4}R^{(2)} + 1\right)\over
\det\left(-\Delta_{cl}+2\right)
\det^{\half}\left(-\Delta_{cl} + 4 - R^{(2)}\right)
\det^{{5\over 2}}\left(-\Delta_{cl}\right)}. }
Recall that $e^{-A_R}$ is the classical contribution, $A_R$ being the 
(regulated) area of the worldsheet; and that the fermionic operator in 
the numerator is a $4\times 4$ matrix operator. The formal
determinants in \effa\ suffer from potential divergences and need to
be regulated. Out of many ways to make sense of these, the heat 
kernel regularization is particularly convenient. There is a vast
literature on this --- we will use Ref.\ \ref\Gilkey{\GILKEY},
which gives an asymptotic expansion for (the logarithm of) such 
determinants as an infinite sum. The first few terms in the sum are 
(regularized) divergent contributions. 

Let $\Lambda$ be an ultraviolet cut-off. It is then easy to see 
(using the results of \Gilkey), that the quadratic divergence 
$c_2\Lambda^2$ cancels between the bosons and fermions. In addition, 
since we have a worldsheet with boundary, there is a linear divergence
$c_1\Lambda$, which cancels in the same way. Finally, the coefficient 
$c_0$ of the logarithmically divergent term $c_0\ln\Lambda$,
is given by the difference in the `mass-square' between bosons and
fermions. We find that, contrary to naive expectation, this coefficient
does not vanish. Specifically, we find that this term is proportional
to
\eqn\logdiv{
c_0\sim
\int d^2\sigma\,\sqrt{h_{cl}}R^{(2)} = 4T\int_{U_0}^\infty dU
{U^4+U_0^4\over U^2\sqrt{U^4-U_0^4}}, }
where we have used the classical solution \usigma. 

Notice that if we were working in the Neveu-Schwarz-Ramond formalism,
divergence of the form \logdiv, (modulo subtleties involving the 
boundary), would have signalled a non-zero $\beta$-function. And to
restore conformal invariance of the sigma model, one will need to
shift the dilaton. In the Green-Schwarz approach that we are working
with, such a conclusion is far from obvious, as there is no
correpondence between the $\beta$-function and equations of motion
of the spacetime fields. 
Therefore one should be cautious of such an interpretation with its 
implication apparently at variance with the conformal invariance of 
the SYM theory. This point is worth further critical 
examination.

While we really do not know the full significance of the logarithmic 
divergence, let us nevertheless try to understand it in our context. 
To this end, we evaluate the coefficient $c_0$ by substituting the upper 
limit of $U$-integration by a cut-off $U_{max}$. This is not a new scale, 
but was already introduced to regularize the classical 
contribution\Malloop. Now we expand \logdiv\ in terms of the small 
parameter $U_0/U_{max}$,
\eqn\logdivtwo{
c_0 \sim
\int d^2\sigma\,\sqrt{h_{cl}}R^{(2)} = 4TU_{max}\left( 1 +
{\cal O}\left({U_0^4\over U_{max}^4}\right)\right). }
To leading order this does not depend on the separation between the
quarks, and goes to only renormalize the (infinite) mass of the quarks.
The higher corrections vanish in the limit $U_{max}\to\infty$. 

We recall that the $\adsf\times\s5$ background was the near horizon
limit of $N$ D3-branes. However, the set up to calculate the quark 
potential differs in one small, but important way. Here one starts 
with $N+1$ D3-branes and takes one of them far away from the others. 
This introduces a scale, whose only effect in the near horizon limit 
is seen in the (infinite) mass of the heavy quarks. The logarithmic 
divergence 
at one-loop goes only to `renormalize' this hidden scale. Therefore 
we conclude, (albeit with some caution), that the one loop 
regularization does not affect the potential energy between the 
quark-antiquark pair.

Some of the higher (non-divergent) terms in the infinite sum in the 
expression of the determinants can be read from Ref.\Gilkey. It would be 
nice to compute the determinants in closed form.

\newsec{Conclusion}
In this letter we study the effect of stringy fluctuations around the
classical macroscopic string background that define expectation value
of Wilson loop operator in the adS/CFT framework. To this end we expand 
the Green-Schwarz type superstring action for the $\adsf\times\s5$ 
background\MetTs\ to second order in fluctuation around the classical 
solution in Ref.\Malloop.  
Both reparametrization as well as the local fermionic $\kappa$-symmetry
is fixed for this background leaving only physical degrees of freedom.
We fix diffeomorphism not by the standard choice of a (conformally) 
flat metric on the worldsheet, but rather by fixing it to be the metric
induced from the target space. Our $\kappa$-symmetry fixing condition
likewise differs from that given in Refs.\KallColl, and is more 
suitable for the problem at hand. We comment on the evaluation of the 
determinants that are the result of functional integration. Surprisingly, 
we find that the divergent contributions do not completely cancel between 
the bose and fermi fields. Our understanding of this is admittedly 
somewhat tentative, and we leave this issue open for further
exploration. 

\bigskip
\leftline{\bf Note added:} After we completed this work, the paper
\ref\GrOl{\GROL}\ appeared in the archive. This also studies stringy
fluctuations affecting Wilson loop in adS/CFT, but in the finite
temperature case. Same applies to \ref\SNaik{\SNAIK}. The techniques
of the present paper have been used in \ref\StFor{\FORSTE} to study 
the fluctuations of membranes of M-theory in $\adss\times\sfour$ 
background. Some preliminary result of the present paper was reported 
in the Ahrenshoop Workshop in Buckow, Germany (September, 1998) by S.F.\ 
and in the String Workshop in Puri, India (December, 1998) by D.G.

\bigskip

\centerline{\bf Acknowledgement}
The work of S.F.\ was supported in part by GIF (German Israeli
Foundation for Scientific Research) and by the EC programmes 
ERB-FMRX-CT-96-0045 and 96-0090. D.G.\ carried out part of the
work at the University of Munich with a fellowship from the Alexander 
von Humboldt Foundation, the hospitality and support of both he
gratefully acknowledges. We would like to thank Jacek Pawelczyk for 
collaboration at the initial stage of this project, and for many useful 
discussions. It is a pleasure to acknowledge fruitful discussions with 
Harald Dorn, Renata Kallosh, Ashoke Sen and Max Zucker. 

\listrefs 
\bye